\documentclass[acus]{jac2000} 
\usepackage{epsfig}
\def\gappeq{\mathrel{ \rlap{\raise.5ex\hbox{$>$}}
                      {\lower.5ex\hbox{$\sim$}}  } }

\def\lappeq{\mathrel{ \rlap{\raise.5ex\hbox{$<$}}
                      {\lower.5ex\hbox{$\sim$}}  } }
\begin{document}
\title{Investigations of Slow Motions of the SLAC Linac Tunnel
\thanks{Work supported by the U.S. Department of Energy, 
Contact Number DE-AC03-76SF00515.}}
\author{Andrei Seryi \\
{\it Stanford Linear Accelerator Center, Stanford University, 
Stanford, California 94309 USA}
}
\maketitle
\begin{abstract}
Investigations of slow transverse motion of the linac tunnel of the Stanford Linear Collider have been performed over period of
about one month in December 1999 -- January 2000. The linac laser alignment system, equipped with a quadrant
photodetector, allowed submicron resolution measurement of the motion of the middle of the linac tunnel with respect to its ends.
Measurements revealed two major sources responsible for the observed relative motion. Variation of the external atmospheric
pressure was found to be the most significant cause of short wavelength transverse motion of the tunnel. The long wavelength component of the motion has been also observed to have a large
contribution from tidal effects. The measured data are
essential for determination of parameters for the Next Linear Collider.
\end{abstract}

\section{Introduction}

The electron-positron linear colliders envisioned for the
future must focus the beams 
to nanometer beam size 
in order to achieve design luminosity.
Small beam sizes impose strict tolerances on the positional 
stability of the collider components, but 
ground motion will continuously 
change the component positions.

For linear colliders, 
the ground motion can be specifically categorized into fast 
and slow motion. Fast ground motion (roughly $f\gappeq 0.1$Hz) causes 
the beam position to change from pulse to pulse. In contrast,
slow ground motion ($f\lappeq 0.1$~Hz) 
does not result in an offset of the beams at the interaction point
since it is corrected by feedback on a 
pulse to pulse basis. However slow motion causes
emittance dilution since 
it causes the beam trajectory to deviate from the ideal line. 
Investigations of slow ground motion are essential to 
determine the requirements for the feedback systems 
and to evaluate the residual emittance dilution 
due to imperfections in the feedback systems.

Investigations of slow motion of 
the SLAC linac tunnel, described in this paper, 
were performed in the framework 
of the Next Linear Collider \cite{ZDR}. 
The measurements were taken from December 8, 1999
to January 7, 2000. Earlier measurements using the same 
technique were performed at SLAC in November 1995 for 
a period of about 48 hours \cite{m95}. 
The goal of the measurements was to 
systematically study the slow motion and to find 
correlations with various external parameters
in order to identify the driving causes.

\section{Results and discussion}

The measurements of slow ground motion 
were performed using the SLAC linac laser 
alignment system \cite{alsys}. This system 
consists of a light source, a detector, and about 300 targets,
one of which is located at each point to be aligned over a total 
length of 3050~m. The targets are installed in a 2-foot diameter 
aluminum pipe which is the basic support girder for
the accelerator. The target is a rectangular Fresnel lens
which has pneumatic actuators that allow 
each lens to be flipped in or out.
The light source is 
a He-Ne laser shining through a pinhole diaphragm.
The beam divergence is large enough to cover even 
nearby targets and only transverse position of the laser, 
but not angle, influences the image position.
The lightpipe is evacuated to about 15 microns of Mercury
to prevent deflection of the alignment image due to refraction in air.
Sections of the lightpipe, which are about 12 meters long,
are connected via bellows that allow independent motion or
adjustment.

\begin{figure}[b]
\vspace{-.65cm}
\hspace{-.3cm}
\centering
{\vbox{
\epsfig{file=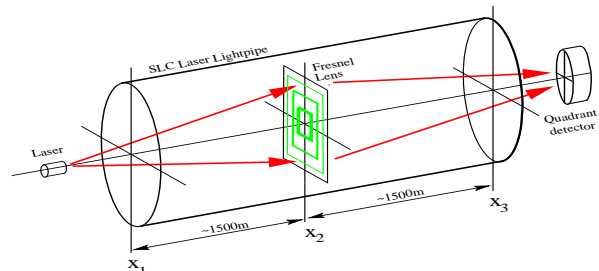,width=0.44\columnwidth,height=0.95\columnwidth,angle=-90}
}}
\vspace{-.3cm}
\caption{Schematic of the measurement setup. }
\vspace{-.07cm}
\label{measscheme}
\end{figure}

A schematic of the measurement setup is shown in Fig.\ref{measscheme}. 
The measurements
were done with a single lens inserted which 
was not moved until the measurements were finished in order 
to ensure maximal accuracy. (In multi target mode 
the repeatability of the target positioning  
limits the accuracy). We used the lens 14-9 located 
at the end of the 14th sector 
of 30 total, 
almost exactly in the middle of the linac.

For these measurements, we replaced the standard detector for this
system with a quadrant
photodetector (produced by Hamamatsu) which has a
quadratic sensitive area ( $\sim 10\times 10$~mm$^2$) 
divided into four sectors.
By combining preamplified signals $u_i$ from these quadrants,
the quantity to be measured $X=x_1+x_3-2x_2$ (see Fig.\ref{measscheme}) 
can be determined as $X \propto [(u_1+u_2)-(u_3+u_4)]/\Sigma u_i$ 
for both the horizontal and vertical ($Y$) planes. 
Calibration of the system was done by moving the detector transversely. 
The sensitivity is linear in the range of $\pm1$~mm.

\begin{figure}[t]
\vspace{-.3cm}
\hspace{-.01cm}
\centering
{\vbox{
\epsfig{file=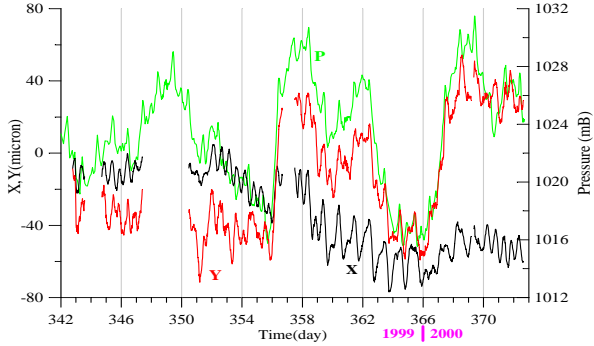,width=0.58\columnwidth,height=0.95\columnwidth,angle=-90}
}}
\vspace{-.4cm}
\caption{Measured horizontal $X$ 
and vertical $Y$ displacements 
plotted along with external atmospheric
pressure.}
\vspace{-.6cm}
\label{alldata}
\end{figure}

The measured data are shown in Fig.\ref{alldata}. 
Two particular characteristics are clearly seen:
the tidal component of the motion is very pronounced and 
there is a strong correlation of the motion 
with external atmospheric pressure.

The linac tunnel was closed, 
with temperature
stabilized water through the RF structures 
during the entire period of the measurements. 
The girder temperature was stable within 0.1$^o$C 
over a day and within a few 0.1$^o$C over a week. 
The RF power was switched off starting Dec.~24 and turned 
on again Jan.~3. This resulted in a slow (weekly) change 
of the girder temperature by 0.5$^o$C in the middle
of the linac 
and 1.5$^o$C at the beginning. 
The average external temperature varied 
by about 10$^o$C over the month. 
No significant correlation of the measured data with 
these and other parameters was observed. 

\begin{figure}[h]
\vspace{-.3cm}
\centering
{\vbox{
\epsfig{file=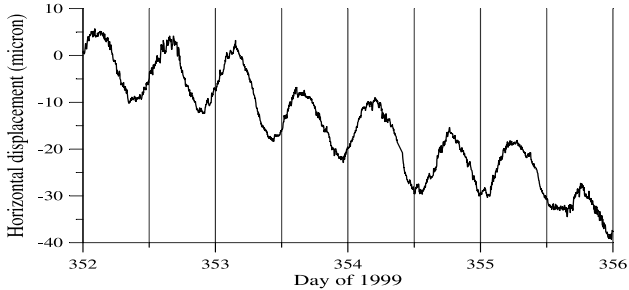,width=0.46\columnwidth,height=1.0\columnwidth,angle=-90}
}}
\vspace{-.35cm}
\caption{Subset of data where tides are seen
most clearly.}
\vspace{-.4cm}
\label{cleartide}
\end{figure}

\begin{figure}[h]
\vspace{-.31cm}
\hspace{-.3cm}
\centering
{\vbox{
\epsfig{file=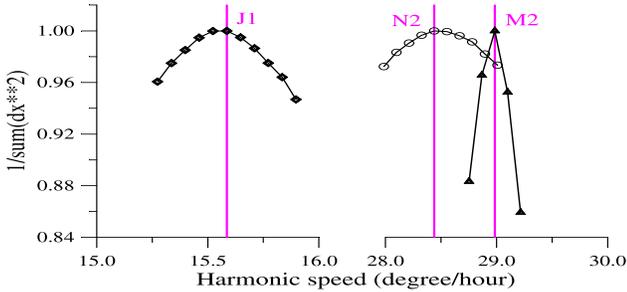,width=0.46\columnwidth,height=1.0\columnwidth,angle=-90}
}}
\vspace{-.2cm}
\caption{Normalized $1/\chi^2$ showing quality of fit
of the measured data by sum of 37 tidal harmonics.
Behavior of $1/\chi^2$ if the speed of one
harmonic would vary. Example for the harmonics
M2 (principal lunar), N2 and J1.
Vertical lines show theoretical speed of these harmonics.}
\label{fit_3h}
\vspace{-.3cm}
\end{figure}

The tidal component of the motion 
has a surprisingly large amplitude ($\sim 10 \mu$m)
(see Fig.\ref{cleartide}). 
The most pronounced harmonics in the measured data 
are M2 (principal lunar), N2 and J1. 
The primary effect of tidal 
deformation is to change the slope 
of the earth's surface ($\sim 100\mu$m/1km
assuming total deformation $\sim 0.5$m).
The secondary effect is to change  
the curvature of the surface ($\sim 0.01\mu$m/1km$^2$
if one assumes uniform earth deformation). 
The laser system is not sensitive to the slope change, 
but only to the curvature change, 
which is an advantage. 
The observed $10 \mu$m change of the curvature 
can only be explained if a local effect
of the tides, with $R_{\mathrm{effective}}\sim 500$km,
is assumed. 
This local anomaly at SLAC is caused by 
loading on the coastline as the ocean water 
level varies due to the tides. This phenomenon has been known
for many years and is called ocean loading. 
This effect is also responsible 
for an enhancement of the tidal variation 
of the earth surface slope observed in 
the San Francisco Bay Area \cite{wood69}.
The ocean loading effect vanishes away from the coastline. 
Regardless, these tidal effects are harmless for a linear collider,
because the motion is slow, very predictable and, 
most importantly, has a wavelength much longer than the 
length of the accelerator. 

\begin{figure}[t]
\vspace{0.2cm}
\centering
{\vbox{
\epsfig{file=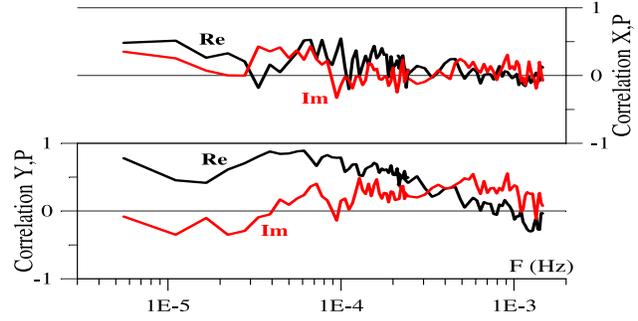,width=0.50\columnwidth,height=1.0\columnwidth,angle=-90}
}}
\vspace{-.2cm}
\caption{Correlation (real and imaginary parts) of displacement 
with atmospheric pressure.}
\vspace{-.5cm}
\label{corrpy}
\end{figure}

Correlation of the tunnel deformation with changes of external 
atmospheric pressure, 
clearly seen in Fig.\ref{alldata} and \ref{corrpy}, 
is significant from the lowest observed frequency up 
to $\sim$0.003Hz. Above this frequency the characteristic 
size over which the pressure changes, which is 
${\sim}v_w/f$,where $v_w$ is the wind velocity
(typically 5m/s), becomes shorter than the linac length
and the correlations vanish. 
In this frequency range, the ratio of deformation to pressure is almost 
constant at about 6$\mu$m/mbar in Y and 2$\mu$m/mbar in X.
The influence of such global changes of pressure 
on the ground deformation can be explained 
if the landscape or the ground properties vary 
along the linac. One should note that 
deformations of the lightpipe itself or motion of the 
targets caused by external pressure variation appear
to be eliminated by design \cite{alsys}.

\begin{figure}[t]
\vspace{.14cm}
\hspace{-.07cm}
\centering
{\vbox{
\epsfig{file=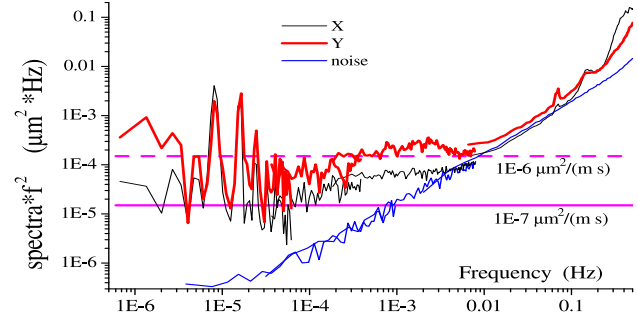,width=0.50\columnwidth,height=1.\columnwidth,angle=90}
}}
\vspace{-0.7cm}
\caption{Spectra of displacement (multiplied by $f^2$) 
and the noise of electronics. Peaks around $10^{-5}$~Hz
correspond to tides. Horizontal lines correspond to the ATL spectra.}
\label{vspec}
\vspace{-.6cm}
\end{figure}

The spectra of the tunnel deformations 
exhibits $1/f^2$ behavior over a large frequency band
(see Fig.\ref{vspec}). 
The $1/f^2$ behavior 
vanishes at $f\gappeq 0.01$~Hz where the signal
to noise ratio becomes poor due to noise 
in the detector and electronics. 
Evaluation of this noise, also shown in Fig.\ref{vspec},
has been done by means of a light source attached
directly to the photodetector. The spot size and intensity 
of this light source were very similar to those 
of the laser. Influences of other sources of error 
(vacuum and temperature variation in the lightpipe,
temperature in the tunnel, etc.) were analyzed but were found 
to be insignificant.

Above 0.1~Hz 
the signal to noice ratio again becomes good as seen in 
the Fig.\ref{vspec}. This is also confirmed by comparison 
of the measured lightpipe displacement with measurements from
a vertical broadband (0.01-100Hz) seismometer STS-2 installed 
at the beginning of the linac, which measures the 
absolute motion of the ground. The simultaneous measurements
of the tunnel motion and of the absolute motion by STS-2
were performed during 3 days from January 4 to January 7.
The coherence between STS-2 and 
vertical displacement measured by photodetector was found 
to be about 0.5 at $F \gappeq 0.2$~Hz. 

\begin{figure}[h]
\vspace{-.64cm}
\hspace{-.07cm}
\centering
{\vbox{
\epsfig{file=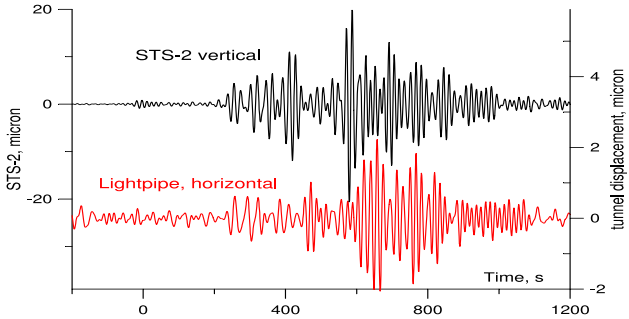,width=0.50\columnwidth,height=1.\columnwidth,angle=-90}
}}
\vspace{-0.3cm}
\caption{Displacement of the tunnel and displacement  
measured by STS-2 seismometer during remote earthquake 
started January 6, 2000 at 02:49:00 local time 
(supposedly corresponds to 5.8MS earthquake at Alaska
happened at 10:42:27 UTC). A passband filter 0.02--0.08Hz
has been applied to the data. }
\label{quake}
\vspace{-.3cm}
\end{figure}

During the 3-day period when the tunnel motion 
was measured simultaneously with STS-2, 
only two remote earthquakes were detected by the seismometer. 
One of the earthquakes did not produce any noticable 
effect on the motion measured by the photodetector, 
probably because of the specific orientation of the waves.
The second earthquake, however, was clearly seen in both 
signals, as shown in Fig.\ref{quake}. The ratio of the
measured absolute motion and the relative deformation 
of the tunnel is consistent with a phase velocity 
of about 2.5~km/s, consistent with earlier 
correlation measurements performed at SLAC \cite{ZDR}.

One model of slow ground motion is described by 
the ATL-law \cite{bb1}. 
For our 3 point motion, the ATL spectrum corresponding 
to the measured $X$ or $Y$ is 
$P(\omega)=\frac{4A L}{\omega^2}$ with $L=1500$m.
Fig.\ref{vspec} shows that the measured spectrum  
corresponds to a parameter $A$ of about 
$10^{-7}$--$2{\cdot}10^{-6}\mu$m/(m$\cdot$s),  
somewhat changing with frequency.

\begin{figure}[t]
\vspace{-1.115cm}
\hspace{-0.018cm}
\centering
{\vbox{
\epsfig{file=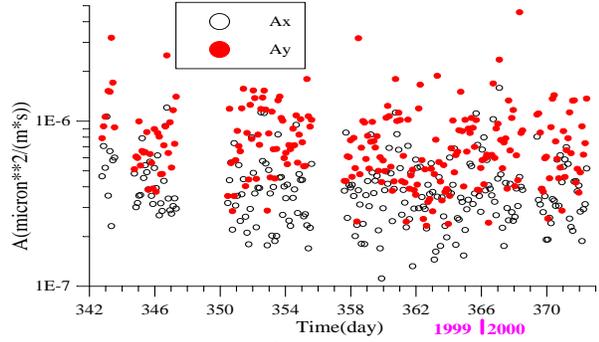,width=0.48\columnwidth,height=0.83\columnwidth,angle=-90,bbllx=10,bblly=100,bburx=430,bbury=686}
}}
\vspace{0.9cm}
\caption{Parameter $A$ defined from fit to spectra in the band
2.44E-4 to 1.53E-2 Hz for all data.}
\vspace{-.1cm}
\label{avst}
\end{figure}

\begin{figure}[t]
\vspace{00.21cm}
\hspace{-0.08cm}
\centering
{\vbox{
\epsfig{file=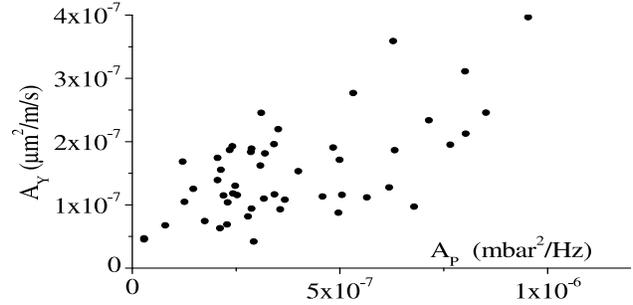,width=0.48\columnwidth,height=1.0\columnwidth,angle=90}
}}
\vspace{-0.8cm}
\caption{Parameter $A_y$ defined from all vertical motion data 
in the frequency band $3\cdot 10^{-5}$--$10^{-3}$~Hz 
versus amplitude $A_p$ of the atmospheric 
pressure spectrum. }
\vspace{-0.6cm}
\label{avsp}
\end{figure}

Spectral analysis of subsets of the data, however, 
shows that this parameter actually varies 
in time (see Fig.\ref{avst}). 
The variation of atmospheric activity 
is again responsible for the variation of parameter $A$.
The spectra of pressure fluctuations was found to behave
also as $A_p/\omega^2$ and its amplitude $A_p$ 
correlates with the parameter $A$, 
as seen in Fig.\ref{avsp}. The temporal pressure variation
can therefore be a major driving term of the 
$A/\omega^2$-like motion. This effect strongly depends 
on geology \cite{as}.


\section*{Conclusion}

Atmospheric pressure changes were found to be a major cause of 
slow motion of the shallow SLAC linac tunnel. 
In deep tunnels or in tunnels built in more solid ground, 
this mechanism would vanish, as it  
and location. Other sources could then dominate.

I would like to thank G.Bowden,
T.King, G.Mazaheri, M.Ross, M.Rogers, L.Griffin, R.Ruland, 
R.Erickson, T.Graul, B.Herrmannsfeldt, R.Pitthan, C.Adolphsen, N.Phinney and 
T.Raubenheimer for help, 
technical assistance and valuable discussions.


\begin{thebibliography}{9}

\bibitem{ZDR} NLC ZDR Design Group, 
SLAC Report-474 (1996). \\[-5mm]

\bibitem{m95}
C.~Adolphsen, G.~Bowden, G.~Mazaheri, in Proc. of LC97. \\[-5mm]


\bibitem{bb1} 
B.Baklakov, P.Lebedev, V.Parkhomchuk, A.Seryi, A.Sleptsov, V.Shiltsev,
Tech.\ Phys.\  {\bf 38}, 894 (1993). \\[-5mm]

\bibitem{alsys}
W.~B.~Herrmannsfeldt,
IEEE Trans.\ Nucl.\ Sci.\  {\bf 12}, 9 (1965).\\[-5mm]

\bibitem{wood69}
Milton D. Wood, Ph.D. thesis, Stanford, May 1969.

\bibitem{fftbwire}
R.~Assmann, C.~Salsberg, C.~Montag, 
SLAC-PUB-7303, in Proceed. of Linac 96, Geneva, (1996).\\[-5mm]

\bibitem{as}
A.~Seryi, EPAC 2000, also in these Proceed. \\[-5mm]


\end{thebibliography}
\end{document}